\documentclass{ws-ijmpa}

\begin{document}

\markboth{O.M. Lecian, G. Montani}
{Exponential Lagrangian for the Gravitational Field and the problem of Vacuum Energy}

%
\catchline{}{}{}{}{}
%

\title{EXPONENTIAL LAGRANGIAN FOR THE GRAVITATIONAL FIELD AND THE PROBLEM OF VACUUM ENERGY
}

\author{ORCHIDEA MARIA LECIAN
}

\address{Dipartimento di Fisica and ICRA, Sapienza, University of Rome,\\Piazzale Aldo Moro, 5- 00185 Roma, Italy
\\
lecian@icra\_it}

\author{GIOVANNI MONTANI}

\address{Dipartimento di Fisica and ICRA, Sapienza, University of Rome,\\Piazzale Aldo Moro, 5- 00185 Roma, Italy\\
ENEA -- C.R. Frascati (Department F.P.N.),\\Via 
Enrico Fermi, 45- 00044, Frascati (Roma), Italy\\
ICRANet -- C. C. Pescara,\\Piazzale della 
Repubblica, 10- 65100, Pescara, Italy\\
montani@icra\_it}

\maketitle

\begin{history}
\received{Day Month Year}
\revised{Day Month Year}
\end{history}

\begin{abstract}
We will analyze two particular features of an exponential gravitational Lagrangian. On the one hand, while this choice of the Lagrangian density allows for two free parameters, only one scale, the cosmological constant, arises as fundamental when the proper Einsteinian limit is to be recovered. On the other hand, the vacuum energy arising from $f(R)$ theories such that $f(0)\neq 0$ needs a cancellation mechanism, by which the present value of the cosmological constant can be recast.  

\end{abstract}

\ccode{PACS numbers: 04.20.Fy  ;  98.80.-k.;x 98.80.Es;
}

\section{Basic statements}
The present value of the cosmological constant \cite{erty} is one of the most tantalizing features of the modern universe: experimental data provide evidence for an accelerating universe, while estimations of the vacuum energy yield indeed the Planckian value, corresponding to $10^{120}$ times the observed numbers. This striking contradiction suggests one to look for a cancellation mechanism, able to reproduce the observed data. The main interesting proposals to interpret the presence of Dark Energy can be divided into two classes \cite{okm}: those theories that make explicitly presence of matter and the other ones, which relay on modifications of the Friedmann dynamics.\\
In fact, the substitution of the Ricci scalar $R$ with a function $f(R)$ in the gravitational action, i.e.
\begin{equation}\label{fdr}
S_{G}=-\frac{c^{3}}{16\pi G}\int d^{4}x \sqrt{-g} f(R),
\end{equation}
leads to the generalized Einstein equations
\begin{equation}\label{ein}
-\frac{1}{2}g_{\mu\nu}f(R)+f'(R)R_{\mu\nu}-\nabla_{\nu}\nabla_{\nu}f'(R)+g_{\mu\nu}\nabla_{\rho}\nabla^{\rho}f'(R)=0,
\end{equation}
which can mimic, under suitable hypotheses, the presence of Dark Energy as due to geometrical contributions. The $0-0$ and the $i-i$ components of (\ref{ein}) for the FRW model can be found by varying (\ref{fdr}), expressed as a function of the lapse $N$ and the scale factor $a$, with respect to these two objects, respectively, using the continuity equation $d(\epsilon a^{3})=-3pa^{2}da$, and then setting $N=1$, in the finite-volume limit.
\section{Exponential Lagrangian density}
The choice of an exponential Lagrangian density \cite{expo} offers the intriguing scenario in which, after modifying the Friedmann dynamics, the presence of matter has to be hypothesized for the Einsteinian limit to be recovered. Furthermore, the Taylor expansion of the exponential Lagrangian density allow us to realize that the geometrical components contain a cosmological term too. An important feature of our model arises when taking a Planckian value for the fundamental parameter of the theory (as requested by the cancellation of the vacuum-energy density). In fact, as far as the Universe leaves the Planckian era and its curvature has a characteristic length much greater than the Planckian one, then the corresponding exponential Lagrangian is expandable in series, reproducing General Relativity (GR) to a high degree of approximation. As a consequence of this natural Einsteinian limit (which is reached in the early history of the Universe), most of the thermal history of the Universe is unaffected by the generalized theory, while the cosmological constant, responsible for the universe acceleration, is the only late-time effect. This model is not aimed at showing that the present Universe acceleration is a consequence of non-Einsteinian dynamics of the gravitational field, but at outlining how it can be recognized from a vacuum-energy cancellation. Such a cancellation must take place in order to deal with an expandable Lagrangian term and must concern the vacuum-energy density as far as we build up the geometrical action only by means of fundamental units.
\subsection{Setting the characteristic length}
The indication from cosmic-microwave-background anisotropies suggests one that the most appropriate characterization for the universe thermo-dynamics be provided by an equation of state of the form $p\sim -\epsilon$ \cite{erty}, i.e. a cosmological term. The appearance of a non-zero cosmological constant indicates that $f(R=0)\neq 0$, while the Einstein-Hilbert (EH) action is a linear term in $R$, with the same sign of the previous one.\\
To deal with $f(R)$ as a series expansion, we would have, in principle, to fix an infinite number of coefficients \cite{plb}. However, in what follows, we address the point of view that only one characteristic length fixes the dynamics.\\
As a consequence of this point of view, we fix the explicit function $f(R)=\lambda e^{\mu R}$ where $\lambda$ and $\mu$ are two constants available for the problem. Comparing the first two terms that come from the expansion of such a function (valid in the region $\mu R\ll 1$ ), with the EH action plus a cosmological term, i.e. $L=-\hbar\left( R+2\Lambda \right)/16 \pi l^{2}_{P}$, we can identify $\lambda=2\Lambda$ and $\mu=1/(2\Lambda)$. As required, our gravitational Lagrangian is fixed by one parameter only, which has to be provided by observational data. 
\subsection{The problem of Vacuum Energy}To investigate the generalized FRW dynamics for this model, we analyze the deSitter regime, where a constant vacuum energy density is taken into account. To this end, we take a cosmic scale factor of the form $a=a_{0} e^{\sigma t}$, where $a_{0}$ and $\sigma$ are two constants. It is easy to recognize that, for such a choice, the Ricci scalar rewrites $R=-12\sigma ^{2}/c^{2}$; hence, according to the equation of state of a cosmological constant, $p=-\epsilon$, the Friedmann equation reads
\begin{equation}\label{sopra}
\epsilon=-\epsilon_{\Lambda}e^{-x}\left(1+\frac{x}{2}\right),\ \ \epsilon_{\Lambda}\equiv\frac{c^{4}\Lambda}{8\pi G},\ \  x\equiv \frac{6\sigma^{2}}{c^{2}\Lambda}\, . 
\end{equation}
The expression above has the surprising feature that the energy density would acquire a negative sign\footnote{This unphysical property is formally removed as soon as we expand the exponential term in correspondence to small values of the dimensionless constant $x$, and, restating the usual Friedmann relation, we get 
\begin{equation}\label{qwert}
\epsilon=\epsilon_{\Lambda}\left( \frac{x}{2}-1\right)\Rightarrow \sigma^{2}=\frac{8\pi G}{3c^{2}}\left( \epsilon+\epsilon_{\Lambda}\right).
\end{equation}
Thus, when the expansion rate of the Universe $\sigma$ is much smaller then the cosmological constant $\Lambda$, we get the usual Friedmann relation between matter and geometry. But, though such a standard relation is apparently reproduced as a low-curvature approximation for $x\ll1$, nevertheless its inconsistency shows up when (\ref{qwert}) is restated as 
\begin{equation}\label{14}
x=2\left(\frac{\epsilon}{\epsilon_{\Lambda}}+1\right).
\end{equation}
We see that, by the expression above, for positive values of $\epsilon$ and $\epsilon_{\Lambda}$, the quantity $x$ has always to be greater than two, in clear contradiction with the hypothesis $x\ll1$, at the ground of the derivation of (\ref{qwert}). Though we are dealing with a surprising behavior, due to the negative ratio $\epsilon/\epsilon_{\Lambda}$, i.e., $\Lambda<0$, however this feature offers an intriguing scenario. In fact,  we will apply relation (\ref{14}) to treat the non-observability of the universe vacuum energy in connection with the present Universe acceleration}.\\
Eq. (\ref{sopra}) admits a special vacuum solution ($\epsilon\equiv 0$), which corresponds to the relation $x\equiv \frac{6\sigma^{2}}{c^{2}\Lambda}=-2$. For the choice of a negative cosmological constant, $\Lambda=-\mid\Lambda\mid$, the equation above provides
\begin{equation}\label{sigmaq}
\sigma^{2}=\frac{c^{2}\mid\Lambda\mid}{3}.
\end{equation}
Thus we see that, in vacuum, our model has the peculiar feature of predicting a deSitter evolution in correspondence to a negative $\Lambda$ value. However, it should be noted that, for this value of $x=-R/(2\Lambda)$, the exponential Lagrangian cannot be expanded, and we deal with the full non-perturbative regime with respect to the Einsteinian gravity. It is just the request to deal with an expandable Lagrangian that leads us to deal with the case $x\ll 1$ and to introduce an external matter field. The exponential term is expandable only if $R/(2\mid\Lambda\mid)\ll1$, but this would imply that the dynamics must contain a cosmological term much greater than the Universe curvature, i.e., an inconsistency which apparently prevents us from recovering the Einstein limit. On the other hand, significant contributions from powers $R/(2\mid\Lambda\mid)\leq1$ would be predicted in a regime where the expansion of the exponential term does not hold.
The vacuum dynamics admits a deSitter phase in correspondence with a certain negative value of $\Lambda$ and the additional presence of matter is observable only if its constant energy density and $\epsilon_{\Lambda}$ have opposite signs. Furthermore, the vacuum solution lives in the non-Einsteinian region ($x\ll1$). The universal features of such a matter contribution and its constant value suggest one to identify it with the vacuum energy. Moreover, the cancellation required to get $x\ll1$ is the natural scenario in which a relic dark energy can be recognized.
The reason why the cancellation proposed between the $\Lambda$ term and the vacuum energy density provides the right order of magnitude of the dark-energy contribution can be recognized in the following fact. By the structure of our model, the relic constant energy density must be a factor $\mathcal{O}(R/(2\mid\Lambda\mid))$ smaller than the dominant contribution $\mathcal{O}(\epsilon_{\Lambda})$. Thus if we take the vacuum energy density close to the Planckian value, then the actual ratio $R/(2\mid\Lambda\mid)$ is of order $\mathcal{O}(10^{-120})$. Such a quantity behaves like $\mathcal{O}(l_{Pl}^{2}/L_{H}^{2})$, where $L_{H}\sim\mathcal{O}(10^{27}cm)$ is the present Hubble radius of the universe. However, it must be remarked that such a consideration holds in the case $\epsilon_{\Lambda}$ and the vacuum energy density are the only contributions. If an additional physical matter field is added, then the relic dark energy contribution is simply constrained to be less than the factor $R/(2\mid\Lambda\mid)$ of the vacuum energy.

\end{document}